\documentclass[
	article,			
	11pt,				
	oneside,			
	a4paper,			
	subsection=TITLE,	
	english,			
	]{abntex2}

\usepackage{float}
\usepackage{subfig}

\usepackage{lipsum}
\usepackage{multicol}
\usepackage{color}
\definecolor{K}{RGB}{10,10,10}     
\definecolor{Gy}{RGB}{127,127,127} 
\definecolor{Bl}{RGB}{63,63,255}   
\definecolor{Gr}{RGB}{63,255,63}   
\definecolor{Rd}{RGB}{255,63,63}   
\definecolor{W}{RGB}{255,255,255}  

\newcommand{\THz}{\mathrm{THz}}
\usepackage{siunitx}

\chardef\us=`\_

\usepackage{cmap}				
\usepackage{lmodern}			
\usepackage[T1]{fontenc}		
\usepackage[utf8]{inputenc}		
\usepackage{nomencl} 			
\usepackage{color}				
\usepackage{graphicx}			
		
\usepackage[alf]{abntex2cite}	

\titulo{Solar Radius at Subterahertz Frequencies and its Relation to Solar Activity}
\autor{F. Menezes\thanks{\url{fabianme17@gmail.com}} \and A. Valio\thanks{\url{avalio@craam.mackenzie.br}}}
\local{Center for Radio Astronomy and Astrophysics (CRAAM),\\ Mackenzie Presbyterian University, S{\~a}o Paulo, Brazil}
\data{\textit{Manuscript accepted by \textbf{Solar Physics} in November 2017}}

\makeatletter
\hypersetup{
		pdftitle={\@title}, 
		pdfauthor={\@author},
    	pdfsubject={Modelo de artigo científico com abnTeX2},
	    pdfcreator={LaTeX with abnTeX2},
		pdfkeywords={abnt}{latex}{abntex}{abntex2}{atigo científico}, 
		colorlinks=true,       		
    	linkcolor=K,          	
    	citecolor=K,        		
    	filecolor=K,      		
		urlcolor=K,
		bookmarksdepth=4
}
\makeatother

\setlrmarginsandblock{4cm}{4cm}{*}
\setulmarginsandblock{4cm}{4cm}{*}
\checkandfixthelayout

\SingleSpacing

\begin{document}
\frenchspacing 
\selectlanguage{english}

\begin{center}
	\textbf{\LARGE \imprimirtitulo\\}
	
	\vspace{.6\baselineskip}
	{\large \imprimirautor\\}
	
	\vspace{.6\baselineskip}
	{\small \texttt{fabianme17@gmail.com}\\ \texttt{avalio@craam.mackenzie.br}\\}
	
	\vspace{.6\baselineskip}
	{\small \imprimirlocal\\}
	
	\vspace{.6\baselineskip}
	{\small \imprimirdata\\}
	
\end{center}

\begin{resumoumacoluna}
	The Sun emits radiation at several wavelengths of the electromagnetic spectrum. In the optical band, the solar radius is 695,700 km and this is what defines the photosphere, the visible surface of the Sun. However, as the altitude increases, the electromagnetic radiation is produced at other frequencies, causing the solar radius to change as function of wavelength. These measurements enable a better understanding of the solar atmosphere and the radius dependence on the solar cycle is a good indicator of the changes that occur in the atmospheric structure. We measure the solar radius at the subterahertz frequencies of 0.212 and 0.405 THz -- \textit{i.e.}, the altitude where these emissions are primarily generated -- and also analyse the radius variation over the 11-year solar activity cycle. For this, we used radio maps of the solar disk for the period between 1999 and 2017, reconstructed from daily scans made by the \textit{Solar Submillimeter-wave Telescope} (SST), installed at El Leoncito Astronomical Complex (CASLEO) in the Argentinean Andes. Our measurements yield radii of \ang{;;966.5} $\pm$ \ang{;;2.8} for 0.2 THz and \ang{;;966.5} $\pm$ \ang{;;2.7} for 0.4 THz. This implies a height of $(5.0 \pm 2.0 \times 10^6)$ m above the photosphere. Furthermore, we also observed strong anti-correlation between the radius variation and the solar activity at both frequencies.
 
 \vspace{\onelineskip}
 
\end{resumoumacoluna}

\textual

\section{Introduction}
\label{intro}
The Sun can be considered as a laboratory, being the closest star and therefore the only one for which highly precise values of mass, radius, luminosity, and activity, among others are known. The solar radius is a very important parameter for the calibration of solar atmospheric models and also for understanding the evolution of the Sun and other stars. As stated by \citeonline{mamajek15} in the International Astronomical Union (IAU) Resolution B3, the nominal solar radius is $R_\odot^N=6.957 (1) \times 10^8$ m, \textit{i. e.}, \ang{;;959.63} and is defined as the distance from the Sun's core to the photosphere which is the solar surface. This is the location where the plasma becomes opaque, equivalent to an optical depth of approximately 2/3 \cite{ostlie06} at visible wavelengths.

The solar radius has been measured for centuries. Recently \citeonline{vaquero16} reported measurements of the optical solar radius from 1773 to 2006 with data from the Royal Observatory of the Spanish Navy. Also, over the last five decades, measurements of the solar radius at many different wavelengths have been made using different techniques. Table \ref{prevtab} lists the values at several radio frequencies showing the increase of the solar radius as the frequency decreases. The dispersion of the values is due to the method used in the radius determination, such as the location of the half value of the disk most common intensity or the position of the inflection point of the intensity nearest to the limb, that can be \ang{;;2} or 1500 km larger at subTHz frequencies.

\begin{table}[ht!]
	\centering
	\caption{Solar radius and altitude values at radio frequencies.}
	\begin{tabular}{lccc}
		\hline
		\textbf{Authors} & \textbf{Frequency} & \textbf{Altitude} & \textbf{Radius} \\
		& (GHz) & ($10^6$ m)   & (arcsec)      \\
		\hline
		\small \citeonline{furst79}       & 3     & $80\pm12$    & $1070\pm17$   \\
		\small \citeonline{furst79}       & 5     & $44\pm6$     & $1020\pm9$    \\
		\small \citeonline{bachurin83}    & 9     & $22\pm1$     & $989\pm2$     \\
		\small \citeonline{furst79}       & 11    & $23\pm4$     & $991\pm5$     \\
		\small \citeonline{bachurin83}    & 13    & $30\pm1$     & $989\pm2$     \\
		\small \citeonline{wrixon70}      & 16    & $22\pm3$     & $990\pm4$     \\
		\small \citeonline{selhorst04}    & 17    & $12.3\pm1.1$ & $976.6\pm1.5$ \\
		\small \citeonline{costa86}       & 22    & $16.0\pm0.6$ & $981.7\pm0.8$ \\
		\small \citeonline{furst79}       & 25    & $14\pm3$     & $979\pm4$     \\
		\small \citeonline{wrixon70}      & 30    & $14\pm3$     & $979\pm4$     \\
		\small \citeonline{pelyushenko83} & 35    & $14\pm2$     & $979\pm3$     \\
		\small \citeonline{costa86}       & 44    & $12.5\pm1.0$ & $978.1\pm1.3$ \\
		\small \citeonline{costa99}       & 48    & $17.4\pm1.4$ & $983.6\pm1.9$ \\
		\small \citeonline{pelyushenko83} & 48    & $9.7\pm2.1$  & $973.1\pm2.9$ \\
		\small \citeonline{coates58}      & 70    & $7\pm3$      & $969\pm5$     \\
		\small \citeonline{kisliakov75}   & 74    & $5\pm3$      & $967\pm4$     \\
		\small \citeonline{swanson73}     & 94    & $9\pm3$      & $972\pm5$     \\
		\small \citeonline{alissan17}     & 100   & $3.2\pm3.3$  & $964.1\pm4.5$ \\
		\small \citeonline{labrum78}      & 100   & $5\pm1$      & $966\pm1$     \\
		\small \citeonline{wannier83}     & 115   & $7.0\pm1.6$  & $969.3\pm1.6$ \\
		\small \citeonline{horne81}       & 231   & $6.2\pm2.0$  & $968.2\pm1.0$ \\
		\small \citeonline{alissan17}     & 239   & $1.1\pm1.8$  & $961.1\pm2.5$ \\
		\hline
	\end{tabular}
	\label{prevtab}
\end{table}

Based on temporal observational series throughout many years, the optical solar radius is not constant and shows slight variations between \ang{;;0.01} and \ang{;;0.50} approximately. Such variations are correlated with the 11-year solar activity cycle. At visible wavelengths for example, \citeonline{ulrich95}, \citeonline{rozelot98}  and \citeonline{emilio00}  found positive correlation in their researches, whereas \citeonline{gilliland81}, \citeonline{wittmann93}, \citeonline{delache93} and \citeonline{laclare96} found an anti-correlation. 

At radio wavelengths, the solar radius variations indicate the different heights where these emissions originate. In other words, we can infer the variation of temperature and density throughout the atmosphere with altitude. 

It is also interesting to investigate the variations of the atmosphere in time throughout the years. This can be done by monitoring the solar radius variation at radio frequencies. Nevertheless, there are not many investigation on this matter. \citeonline{bachurin83} reported variations in time of the solar radius at microwaves. In July 1980 and January 1981, the measured radii were $1.031 \pm 0.002 \; R_\odot$ at 13 GHz and $1.043 \pm 0.004 \; R_\odot$ at 8 GHz, that meant an increase of nearly \ang{;;9.6} and  \ang{;;13.8} respectively in comparison with their values in 1976. Another study performed by \citeonline{costa99} at 48 GHz  reported a decrease in radius correlated with the solar cycle between 1991 and 1993. Finally, \citeonline{selhorst04} reported radius variations at 17 GHz from 1992 to 2003. This study revealed correlation with the solar cycle, but when using only measurements from polar latitudes, the radius variation become anti-correlated to the solar activity cycle.

Here we measured the solar radii at subterahertz frequencies of 0.212 and 0.405 THz throughout 18 years, from 1999 to 2017. In Section \ref{method} we explain the observational methods and data analysis. We present the results and discussions in Section \ref{result}. Finally, in Section \ref{conclusion}, we list our conclusions.

\section{Observations and Data}
\label{method}
	\subsection{SOLAR SUBMILLIMETER-WAVE TELESCOPE}
\begin{figure}[ht!]
			\captionsetup{type=figure}
			\centering
			\includegraphics[height=.5\textwidth]{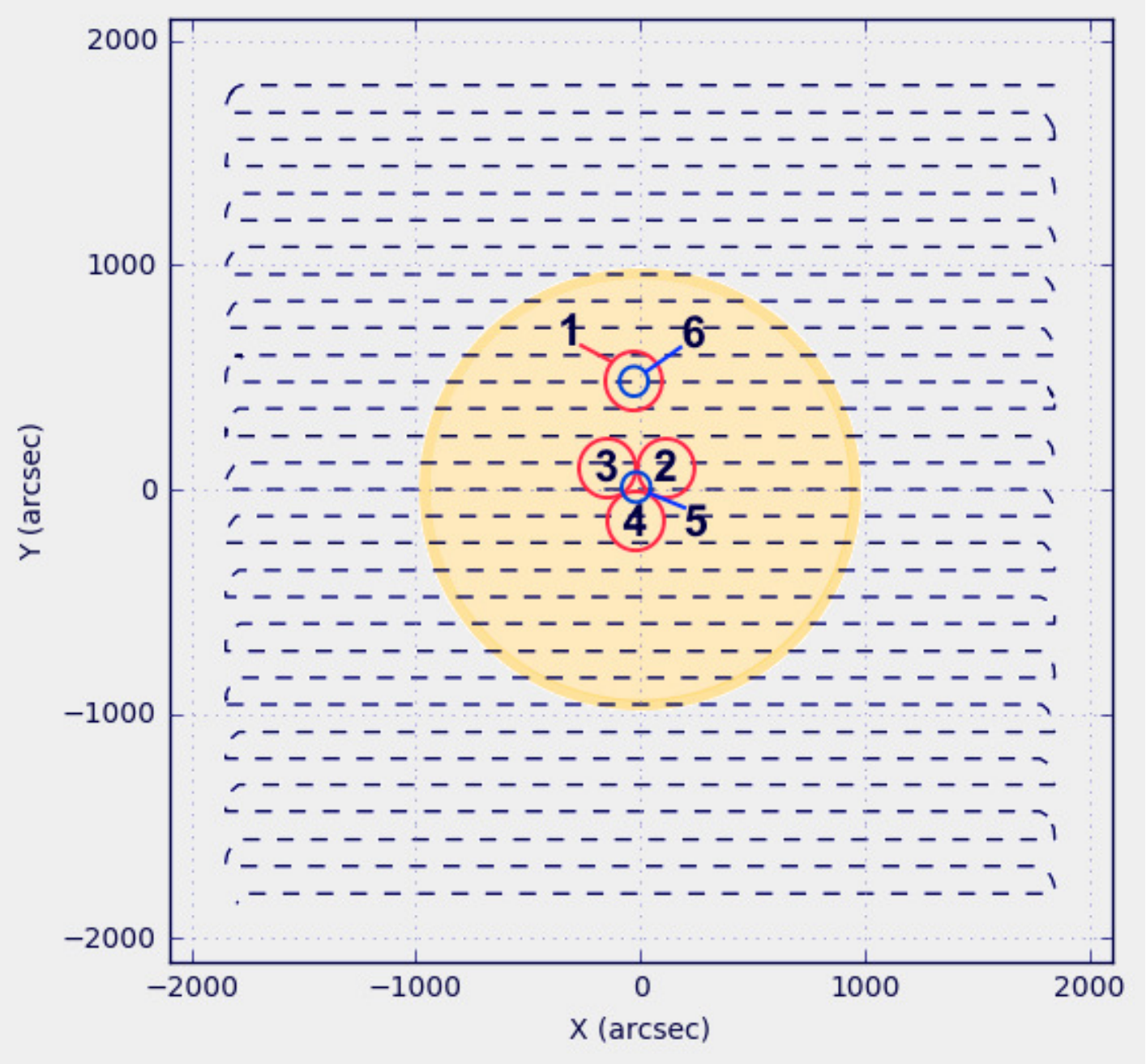}
			\caption{\small Scans (dashed lines) over the solar disk (yellow) with circles representing the six telescope beams.}
			\label{SSTbeams}
		\end{figure}

		The \textit{Solar Submillimeter-wave Telescope} (SST) was the first instrument conceived to continuously study the submillimeter spectrum of solar emission in quiet, quiescent, and explosive conditions \cite{kaufmann01}. The instrument is located in CASLEO, at 2550 m altitude, in the Province of San Juan, Argentina. It has two radiometers at 0.740 mm (405 GHz) and four at 1.415 mm (212 GHz) and their half-power beam width (HPBW) are \ang{;2;} and \ang{;4;} (arcminutes), respectively \cite{kaufmann08}, shown as red circles in Figure \ref{SSTbeams}.
		
\begin{figure}[ht!]
			\captionsetup{type=figure}
			\centering
			\includegraphics[height=.5\textwidth]{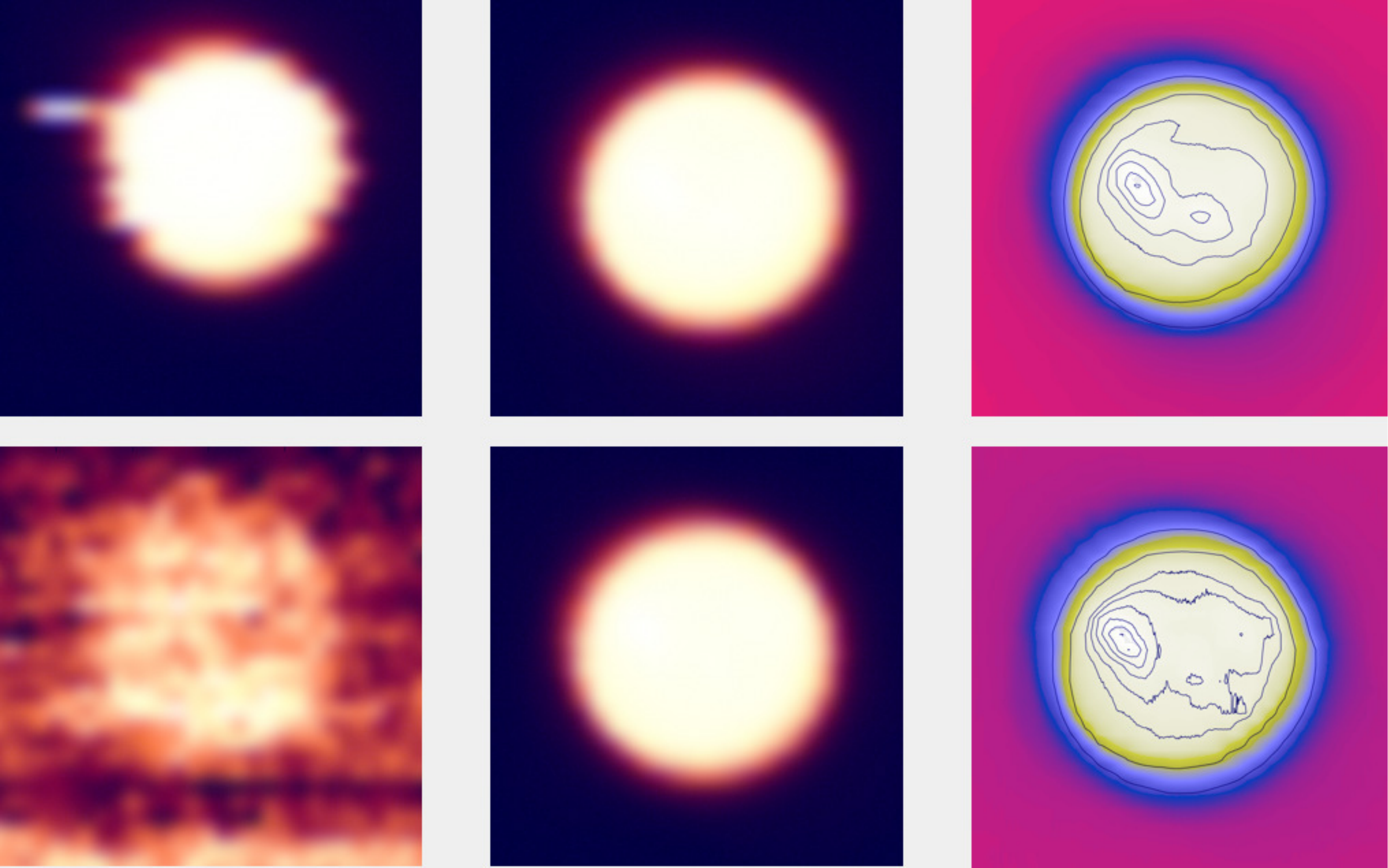}
			\caption{\small Reconstructed solar map examples. \textit{Top-left}: 0.212 THz map from 2007-07-17 with instrumental error. \textit{Bottom-left}: 0.405 THz map from 2008-12-08 with Earth's atmospheric interference (high opacity). \textit{Top-center}: very good 0.212 THz map from 2008-01-09. \textit{Bottom-center}: very good 0.405 THz map from 2008-01-09.  \textit{Top-right}: 0.212 THz map from 2008-01-09 with contour lines of relative intensity levels of 0.5, 0.9, 1.02, 1.04, 1.06, 1.08, 1.095 with respect to the quiet Sun value. \textit{Bottom-right}: 0.405 THz map from 2008-01-09 with contour lines of quiet Sun relative intensity levels of 0.5, 0.9, 1.02, 1.04, 1.06, 1.08, 1.10, 1.12.}
		\label{maps}
	\end{figure}

		This radio-telescope makes solar maps from azimuthal, elevation, right ascension, declination or radial raster scans, however azimuthal scans are the most common, and these were the ones used here. This operation mode has a sampling step of \ang{;;20} in azimuth taken at every 40 milliseconds. The parallel scans, shown as dashed lines in Figure \ref{SSTbeams}, have a separation of \ang{;2;} and are used to reconstruct a 2D image of the Sun. From 1999 to 2017, approximately 3 solar scans per day were performed during 4020 days  which resulted in 11 201 maps for each channel (44 804 for the four 212 GHz channels and 22 402 for the two 405 GHz ones). However, due to mostly high atmospheric opacity (Figure \ref{maps}--left-bottom) or even instrumental errors (Figure \ref{maps}--left-top), many of the maps had to be discarded. Thus remaining 16 623 maps
		in total: 13 200 (29.5\% of initial value) at 212 GHz and 3423 (15.3\%) at 405 GHz. Examples of the discarded maps are shown in the left panels, whereas the good maps are shown in the center panels of Figure \ref{maps}. These are repeated in the right panels of Figure \ref{maps}, with a logarithmic color scale and with contour lines of relative intensity levels to outline the active regions at both frequencies.
		
	\subsection{RADIUS DETERMINATION}
\begin{figure}[ht!]
			\captionsetup{type=figure}
			\centering
			\subfloat
			{\includegraphics[ width=.33\textwidth, height=.285\textwidth ]{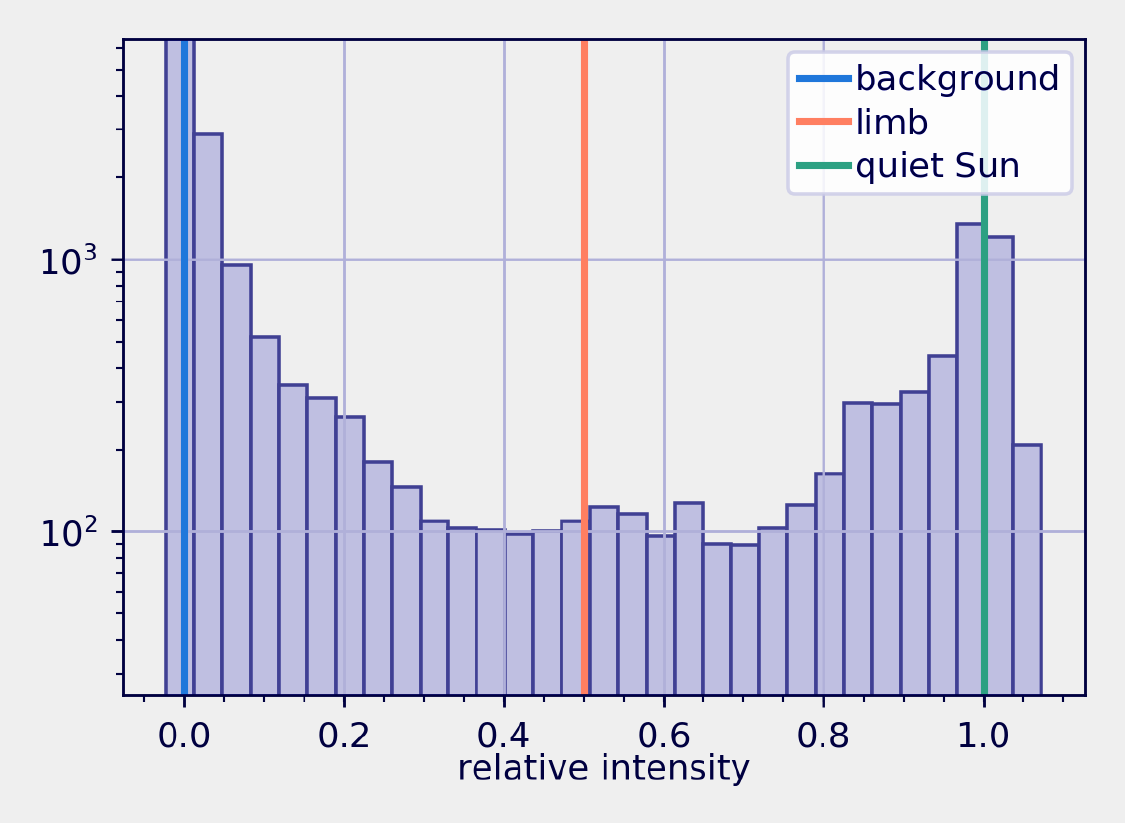}}\hfill
			\subfloat
			{\includegraphics[ width=.33\textwidth, height=.285\textwidth ]{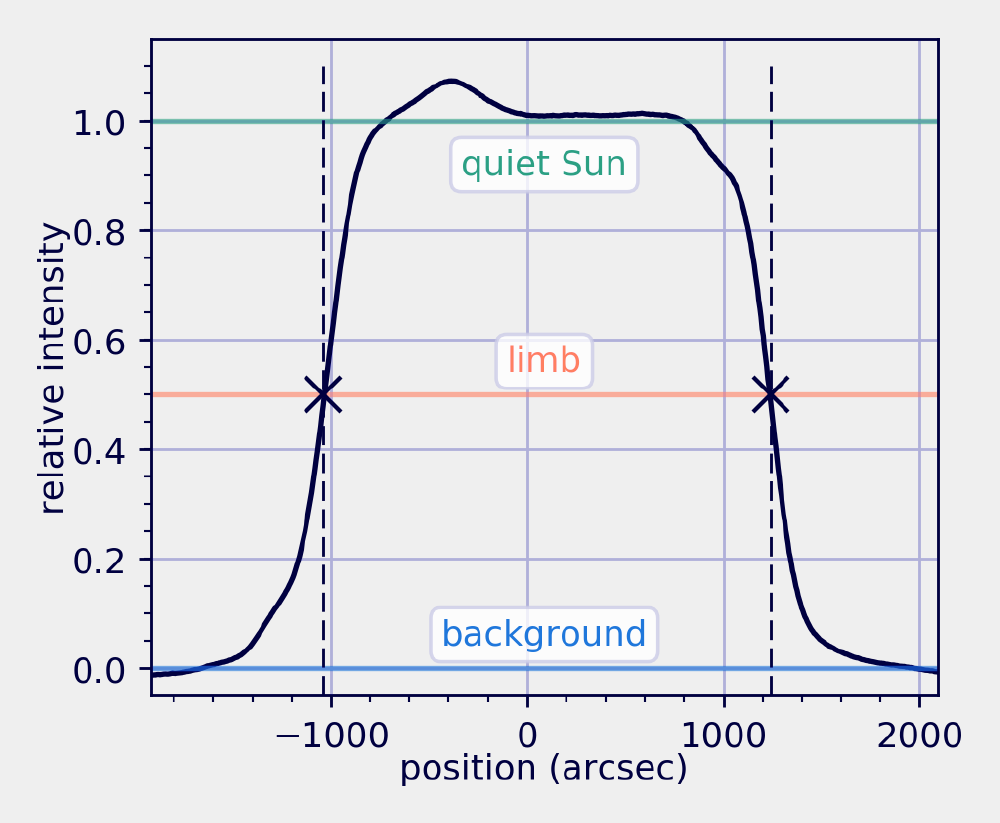}}\hfill
			\subfloat
			{\includegraphics[ width=.33\textwidth, height=.285\textwidth ]{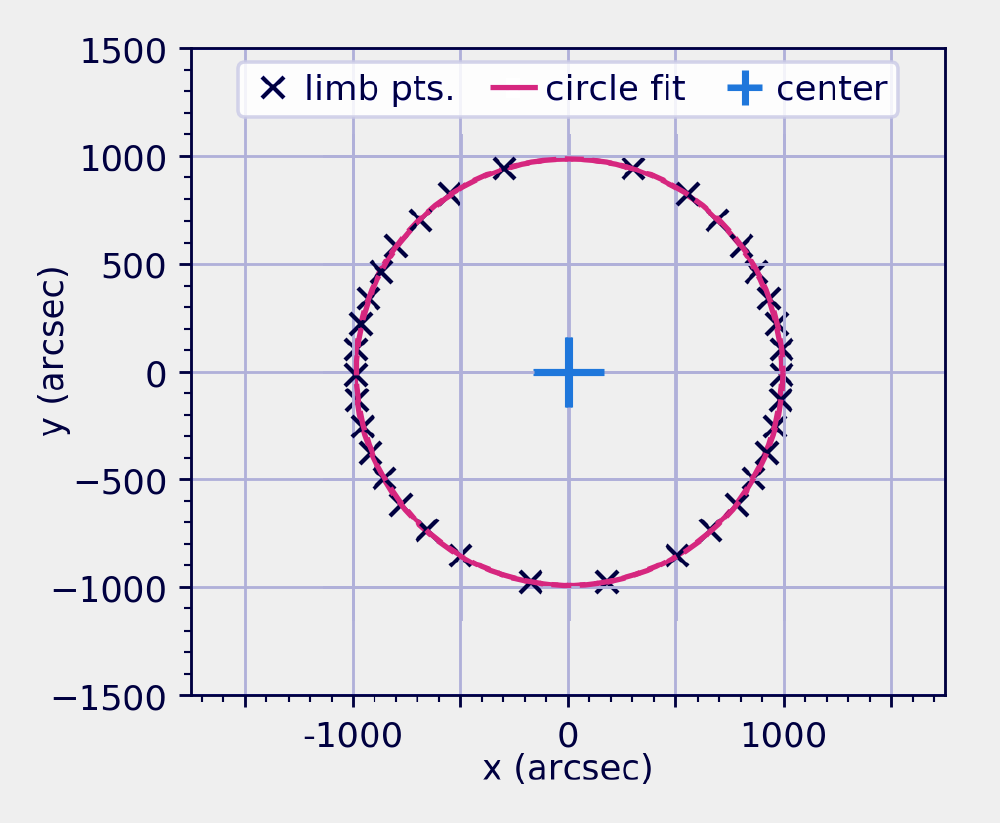}}
			\caption{\small Stages of solar radius determination \textit{Left}: Distribution of the temperature values where the background value (blue line) and quiet Sun level (green line) are defined. \textit{Center}: Indication of the points corresponding to the solar limb of one scan. \textit{Right}: Limb coordinates with circle fit.
			}
			\label{fitmap}
		\end{figure}

		To determine the radius of the Sun it is necessary to define the solar limb. This is done by interpolating the mid value between the background and the quiet Sun levels. The background value is set as the most common temperature (ADC)\footnote{ADC: analog-to-digital converter units that are linearly proportional to temperature.} value in the map as shown in the map intensity histogram (blue line) in Figure \ref{fitmap}--left. The quiet Sun level  is the most common value in the distribution of the solar disk intensity (second peak to the right of the histogram, shown as the green line in Figure \ref{fitmap}--left). Then, the solar limb is interpolated as the point where the intensity is the mean between the quiet Sun and the background levels (Figure \ref{fitmap}--center) for each scan of the map. In Figure \ref{fitmap}--right, the coordinates corresponding to the limb points (black crosses) are then fit by a circle (red line) using a least squares method to determine the center coordinates (blue cross) given by $x_c$ and $y_c$, and the radius:
		\begin{equation}
		R = \sqrt{(x-x_c)^2+(y-y_c)^2}
		\end{equation}
		
		However, as it was mentioned, there are many instrumental errors and errors due to atmospheric attenuation. Thus during the extraction of limb points, four criteria were added to extract the points more precisely. As a first filter, only limb points with distance to the map center (estimated as the average of the coordinates $x$ and $y$ corresponding to intensities above the limb) between $0.985R_{ap}$ (apparent radius) and $1.015R_{ap}$ passed onto the next step. The second filter is applied to the circle fitting step. After a first fit, we discard the points which distance from map center (now obtained by the fitting) is not between $0.995R_{fit1}$ and $1.005R_{fit1}$. In the third filter, a further fitting is made with the points that passed through the previous filter, and now we disregard the points which distance to map center (obtained in this fitting) is not between $1R_{ap}$ and $1,001R_{fit2}$. Then a final fitting is made with the remaining points and the average radius of the map calculated. If, after all these selection stpdf, only 10 points were left, the whole map was discarded due to the low number of limb points.
		
		The resulting radius values were also corrected for the Earth's orbit eccentricity, which makes the apparent radius of the Sun vary between \ang{;;975.3} and \ang{;;943.2} during the year -- greater in perihelion, smaller in aphelion. Next, the visible solar radius was subtracted from the mean sub-THz ones to determine the altitudes where the 0.212 and 0.405 THz emissions are predominantly produced.
		
		To do the analysis of radius dependence on the solar activity cycle we applied a running mean every 1 month and every 13 months, and then used the Pearson's correlation coefficient (PCC) to check if there were any correlation. The proxy used for the solar activity cycle was the sunspot index number (SILSO data, Royal Observatory of Belgium, Brussels).
		
\section{Results and Discussion}
\label{result}
	\subsection{SUBTERAHERTZ RADIUS}
\begin{figure}[ht!]
			\captionsetup{type=figure}
			\centering
			\subfloat
			{\includegraphics[width=.5\textwidth]{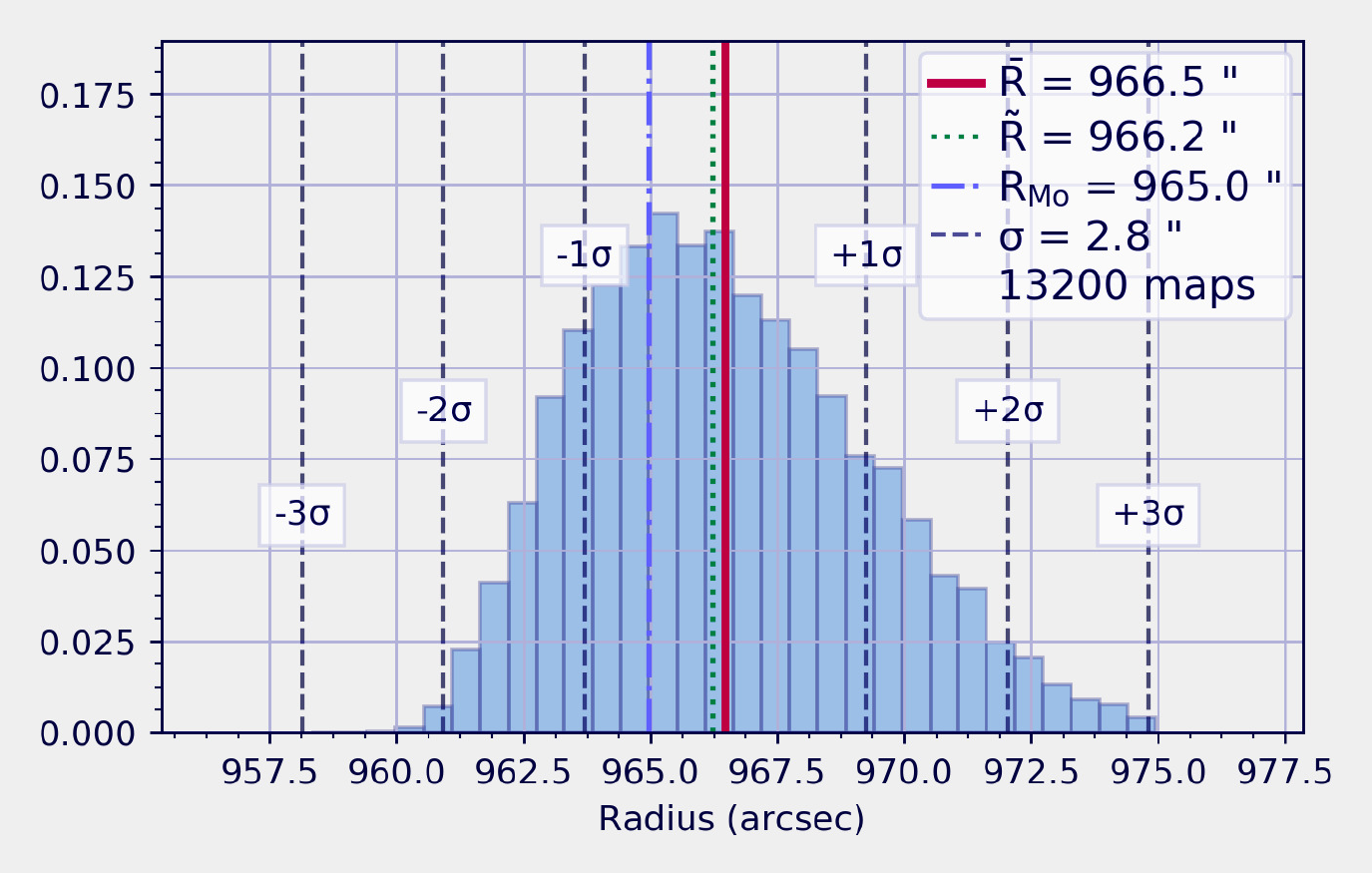}}
			\subfloat
			{\includegraphics[width=.5\textwidth]{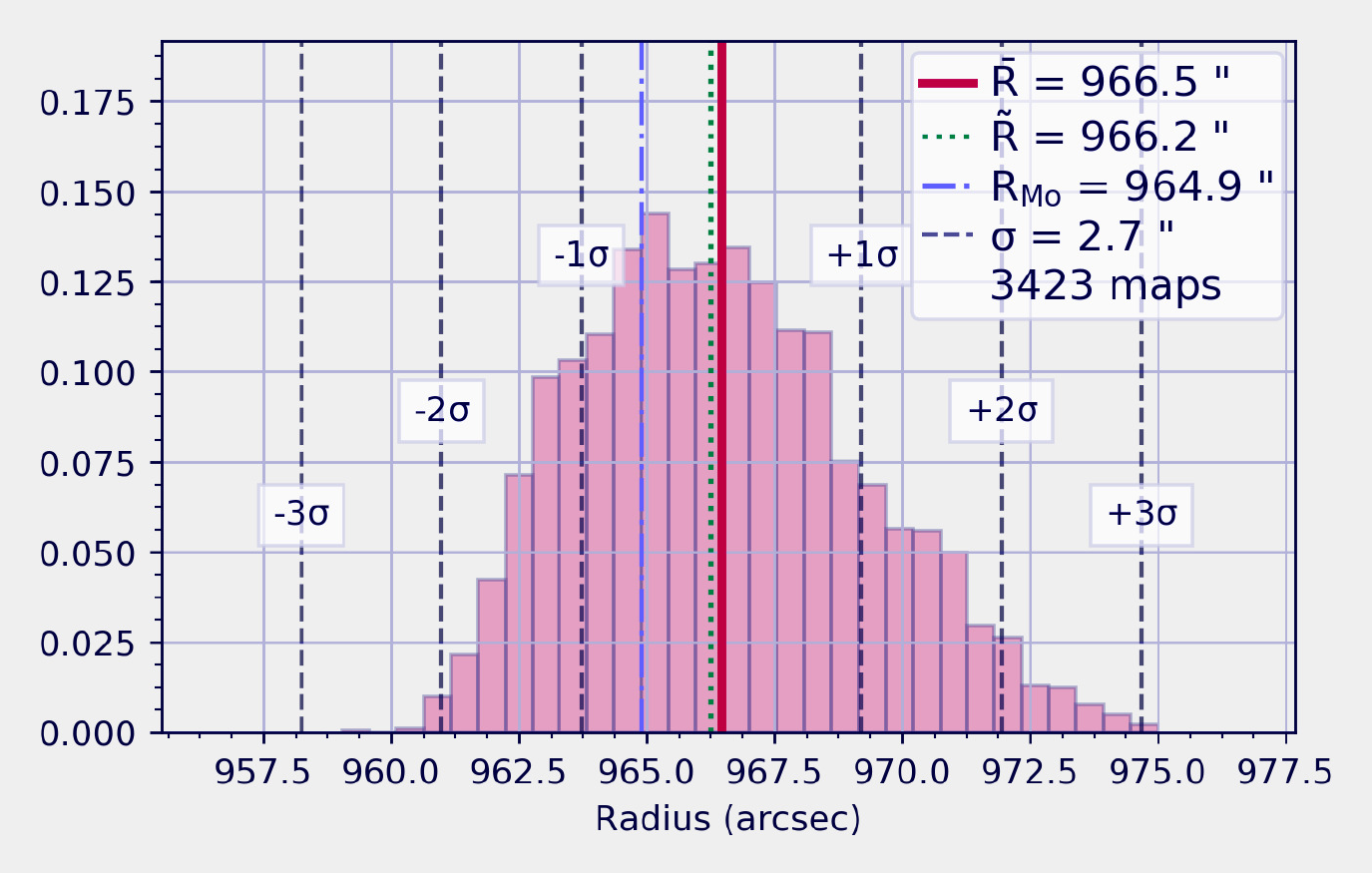}}
			\caption{\small Radius distribution for all maps from 1999 to 2017. \textit{Left}: 0.212 THz channels. \textit{Right}: 0.405 THz channels.}
			\label{histogram}
		\end{figure}

		After the analysis of over 16 600 maps by the method described in Section~\ref{method}, the mean radius obtained for both 0.212 and 0.405 THz was \ang{;;966.5} with rms of \ang{;;2.8} and \ang{;;2.7}, respectively, as indicated in the histograms of Figure \ref{histogram} by red lines. 
		These results are also summarized in Table~\ref{results}. 
		
		\begin{table}[ht!]
			\centering
			\caption{Mean sub-THz and optical radius and altitude values.}
			\begin{tabular}{lcccc}
				\hline
				\textbf{Frequency} & \textbf{Radius}  & \textbf{Radius}      & \textbf{Radius}    & \textbf{Altitude}\\
				& (arcsec)      & ($R_\odot$)     & ($10^8$ m)    & ($10^6$ m)  \\
				\hline
				{212 GHz} & $966.5\pm2.8$ & $1.007\pm0.003$ & $7.01\pm0.02$ & $5.0\pm2.0$ \\
				{405 GHz} & $966.5\pm2.7$ & $1.007\pm0.003$ & $7.01\pm0.02$ & $5.0\pm2.0$ \\
				{Optical} & $959.63$      & $1$             & $6.957$       & $0$         \\
				\hline
			\end{tabular}
			\label{results}
		\end{table}
		
\begin{figure}[ht!]
			\captionsetup{type=figure}
			\centering
			\includegraphics[ height=70mm, width=.8\textwidth ]{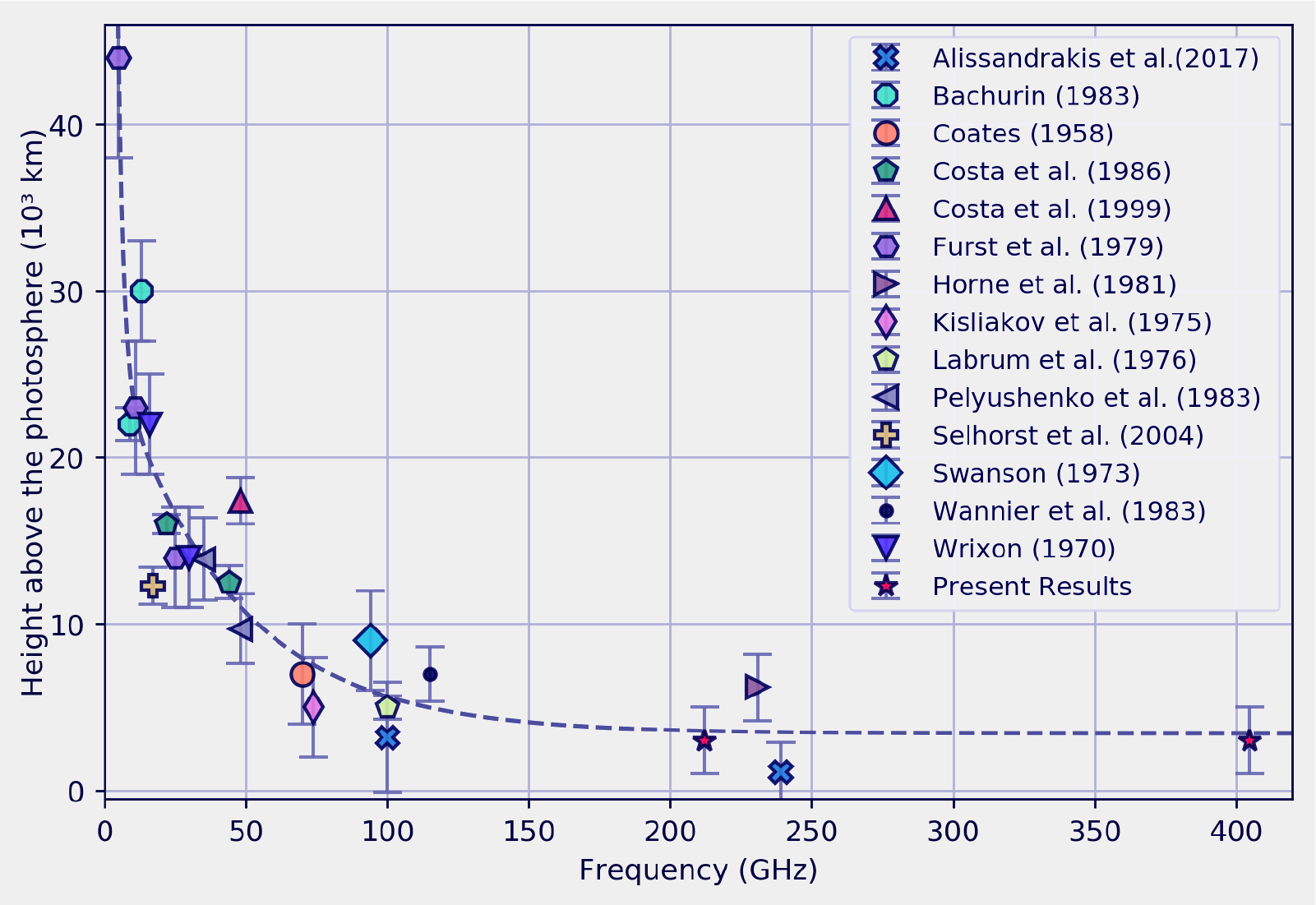}
			\caption{\small Previous altitude values as function of the frequency from other authors and including our present results. The dashed line represents the curve fit to the points (see Equation \ref{eq:exp}).}
			\label{prevgraph}
		\end{figure}

		To check for consistency, we plotted our results with those from other authors (Table~\ref{prevtab}) on Figure \ref{prevgraph}. A double-exponential curve was fit to the points, since the altitude of the emission seemed to be decreasing exponentially with frequency:
		\begin{equation}
		H(\nu)=A \; e^{-B \; \nu} + C \; e^{-D \; \nu} + E \; ,
		\label{eq:exp}
		\end{equation}
		where $H$ is the height above the photosphere in km, $\nu$ is the frequency in GHz and the fit parameters are $A = 2.64\times10^5$ km, $B = 5.27\times10^{-1}$ GHz$^{-1}$, $C = 2.38\times10^4$ km, $D = 2.38\times10^{-2}$ GHz$^{-1}$ and $E = 3.43\times10^3$ km. Our results, shown as red stars in the plot, seem to be in good agreement with the previous ones.
		
\begin{figure}[ht!]
			\captionsetup{type=figure}
			\centering
			\subfloat
			{\includegraphics[ trim={2mm 4mm 0mm 1mm}, clip, width=.35\textwidth ]{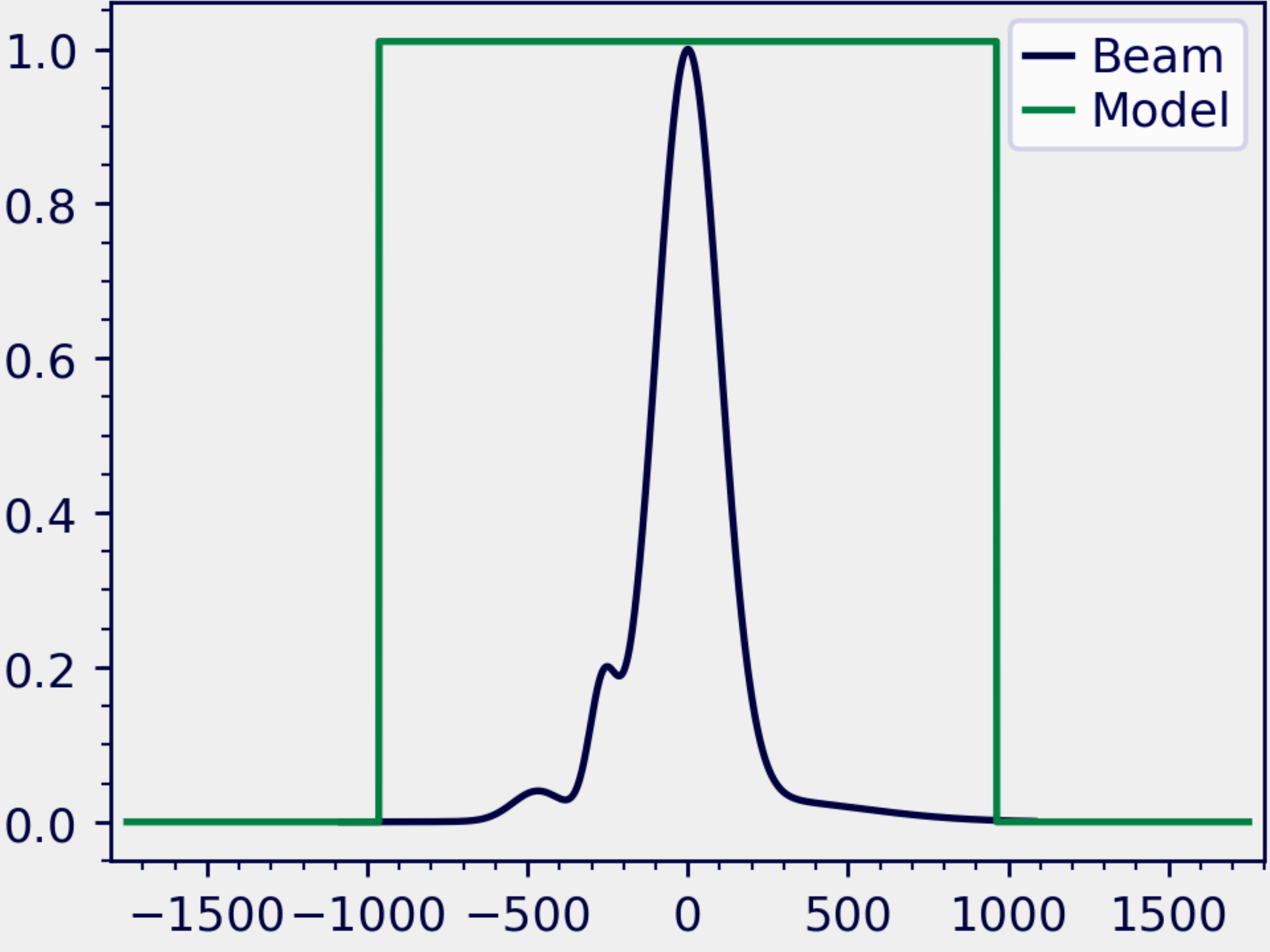}}\hspace{5pt}
			\subfloat
			{\includegraphics[ trim={2mm 4mm 0mm 1mm}, clip, width=.35\textwidth ]{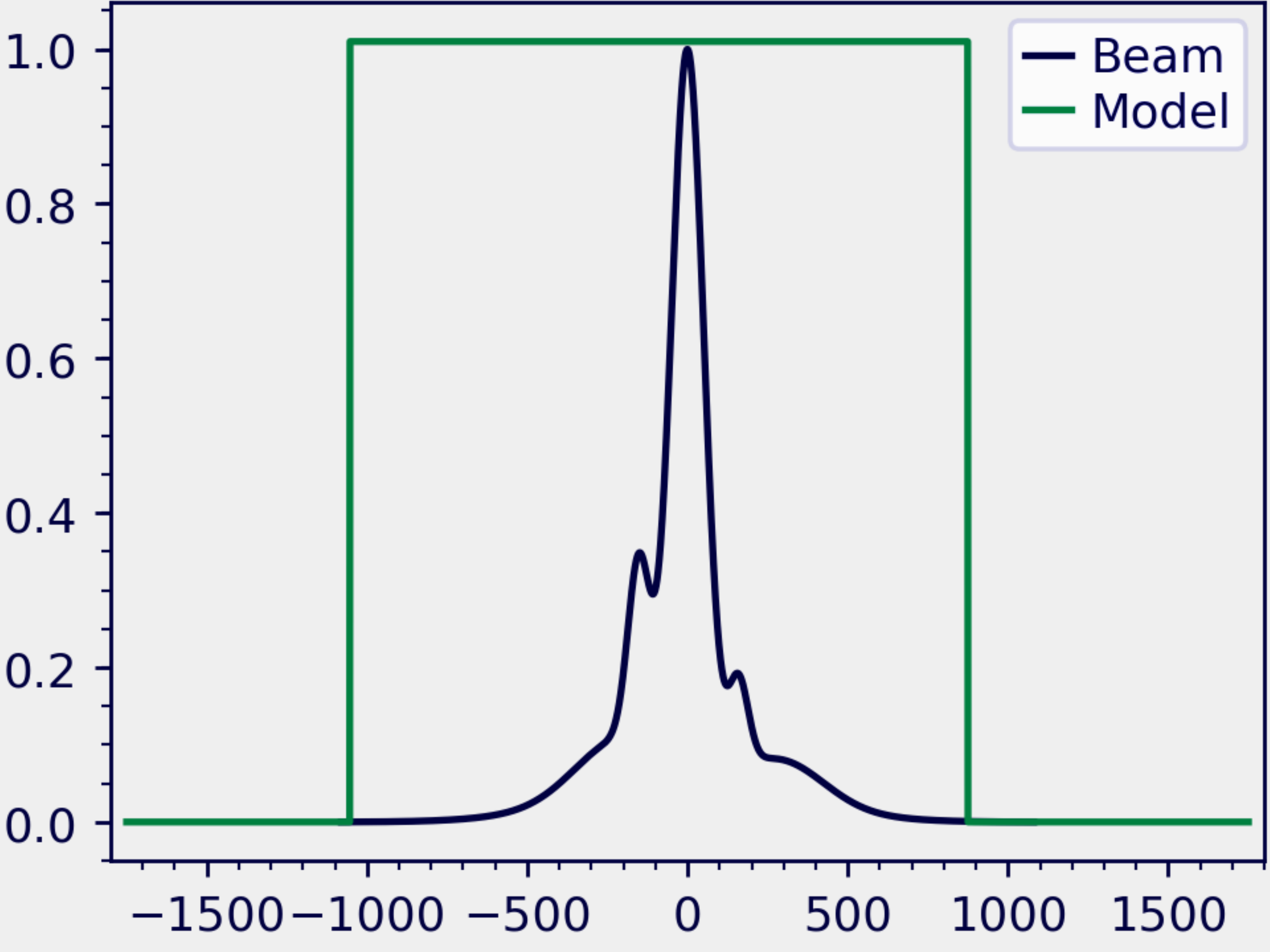}}\quad
			\subfloat
			{\includegraphics[ trim={2mm 4mm 0mm 1mm}, clip, width=.35\textwidth ]{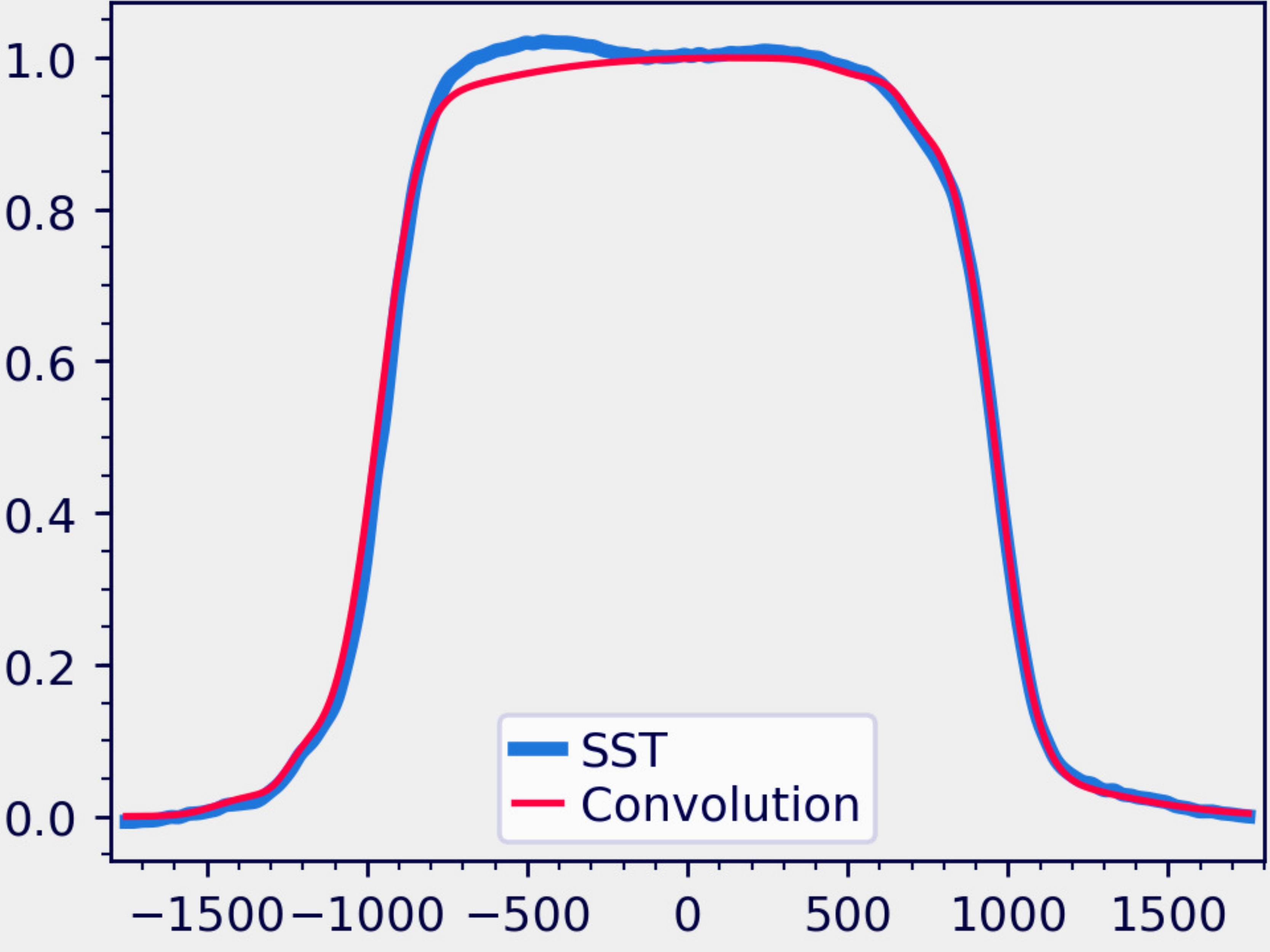}}\hspace{5pt}
			\subfloat
			{\includegraphics[ trim={2mm 4mm 0mm 1mm}, clip, width=.35\textwidth ]{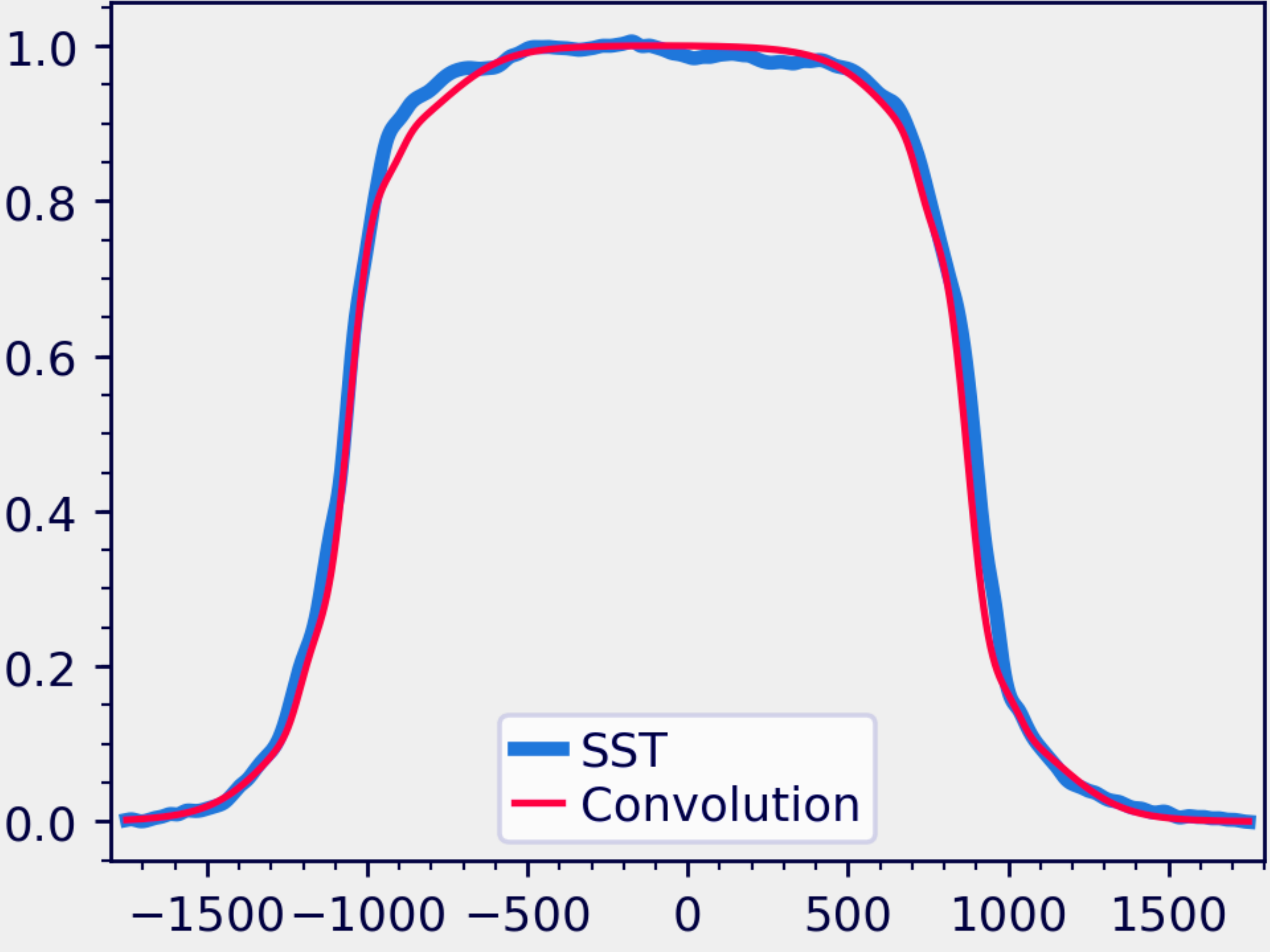}}
			\caption{\small Model Sun brightness temperature profile. \textit{Top}: model cylinder (green) and reconstructed beam (black) for 0.212 (left) and 0.405 GHz (right). \textit{Bottom}: convoluted model (red) and SST observed signal (blue) at 0.212 (left) and 0.405 GHz (right).}
			\label{model}
		\end{figure}

		Another way to check our results was to compare them to a simple model of the Sun. Considering a constant intensity over the whole solar disk -- \textit{i.e.}, a cylinder with a radius of \ang{;;966}, we convoluted this model with the reconstructed SST beams. 
		Next, we applied the same methodology to obtain the limb points at half value of the most common disk intensity and fit a circle to these points.
		For 0.212 and 0.405 THz frequencies, the model yields radii values of \ang{;;966.20} and \ang{;;966.18} respectively, which match our results. Two examples of a model solar scans at these frequencies are shown in Figure \ref{model}.
		
	\subsection{RADIUS TEMPORAL VARIATION OVER THE SOLAR ACTIVITY CYCLE}
		In addition to the solar average radius, we investigated the solar radius variation at subterahertz frequencies with the 11-year solar activity cycle and its relationship with the sunspot index number (SILSO data, Royal Observatory of Belgium, Brussels). At first, for a 1-month running mean applied to the measured solar radius at THz frequencies, we obtained $-0.341$ for 0.212 THz and $-0.227$ for 0.405 THz as correlation coefficients with the sunspot index number, respectively, which indicates barely no correlation at all. Next, we tried a longer running mean of 13-month, and the correlation coefficients increased to $\rho_{0.212 \THz} = -0.521$ and $\rho_{0.405 \THz} = -0.675$, which are moderate coefficients.
		
\begin{figure}[ht!]
			\captionsetup{type=figure}
			\centering
			\subfloat
			{\includegraphics[trim={2mm 3mm 2mm 3mm}, clip, height=35mm, width=.5\textwidth ]{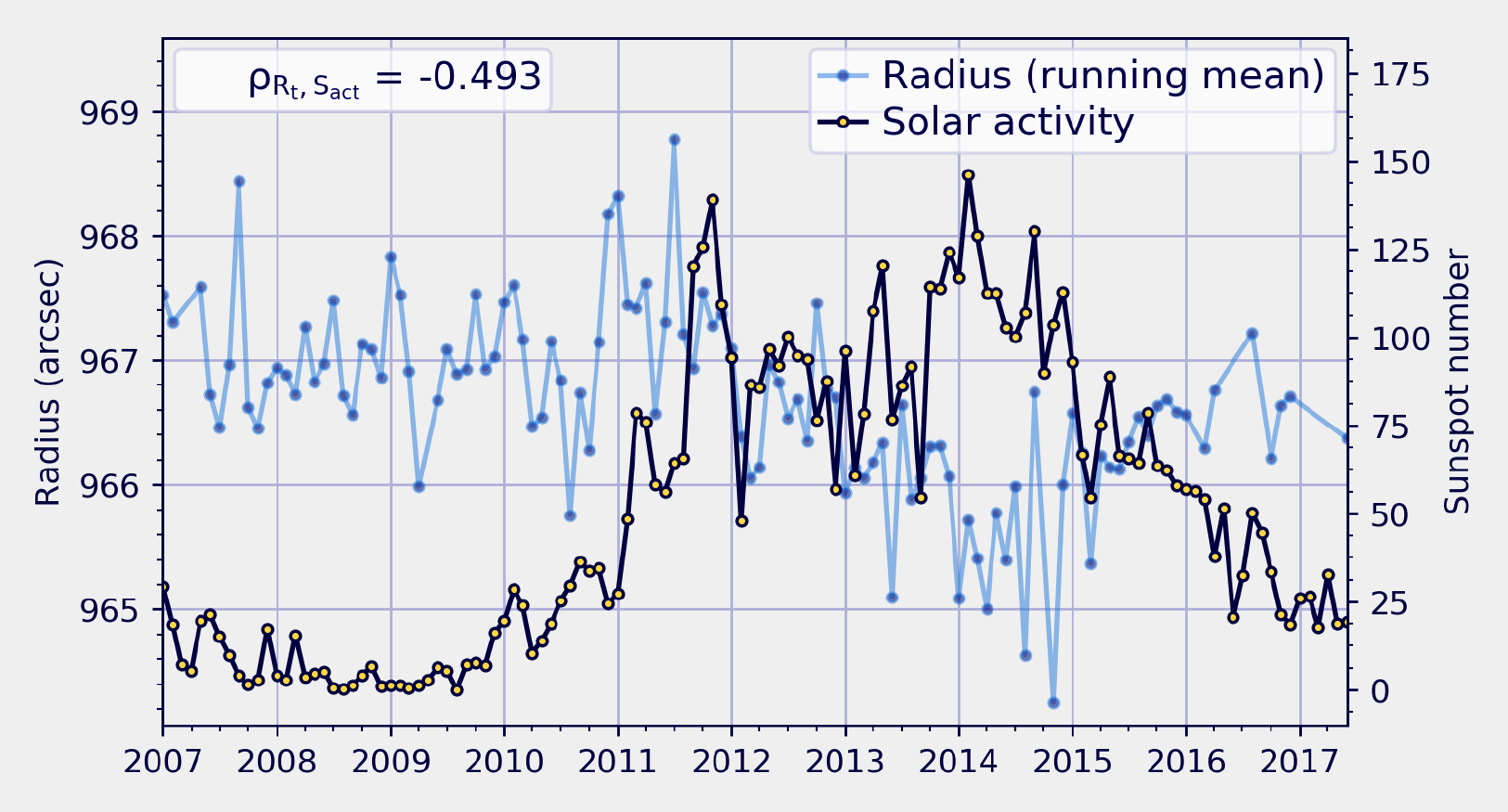}}
			\subfloat
			{\includegraphics[trim={2mm 3mm 2mm 3mm}, clip, height=35mm, width=.5\textwidth ]{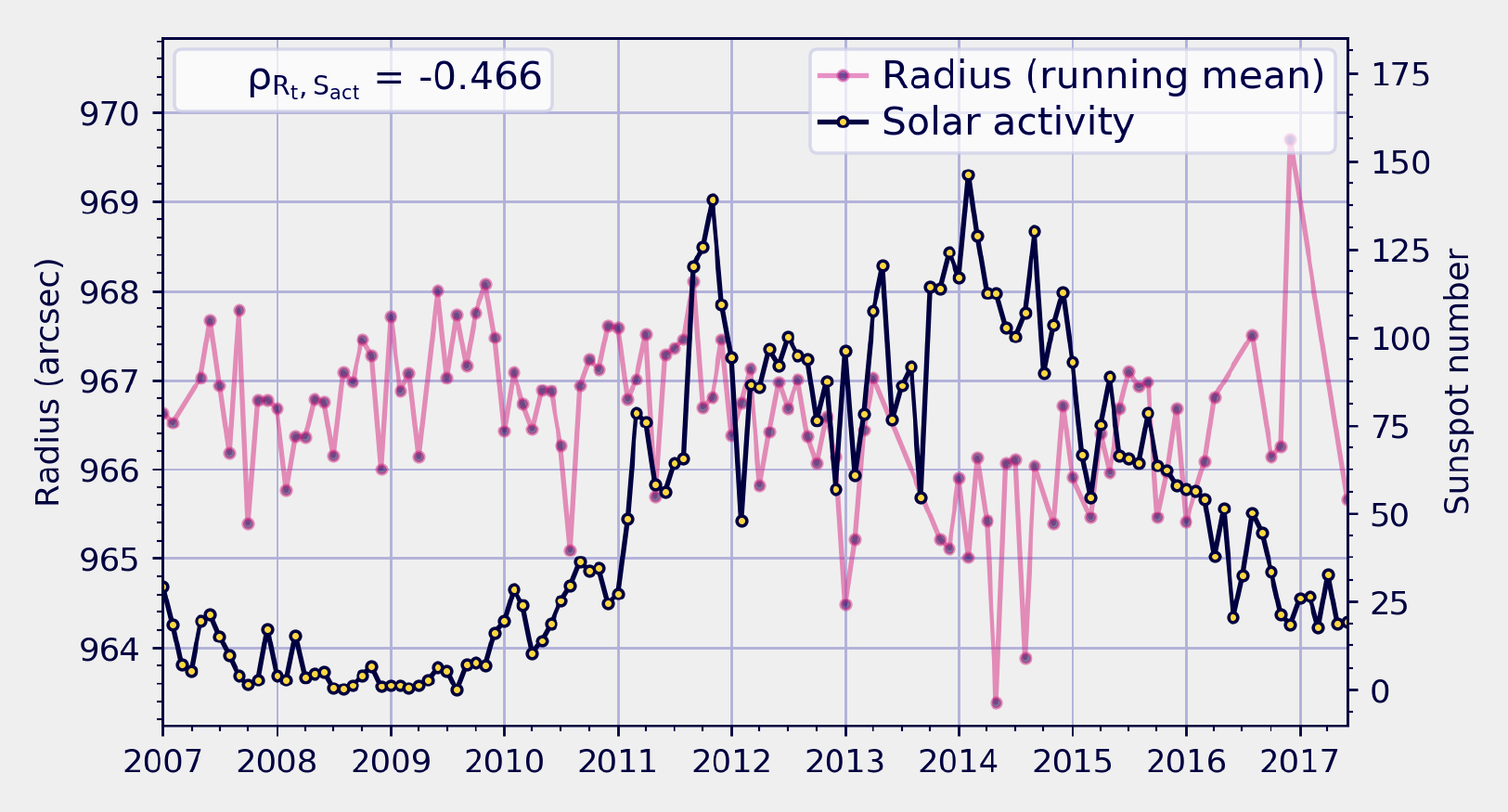}}\quad
			\subfloat
			{\includegraphics[trim={2mm 3mm 2mm 3mm}, clip, height=35mm, width=.5\textwidth ]{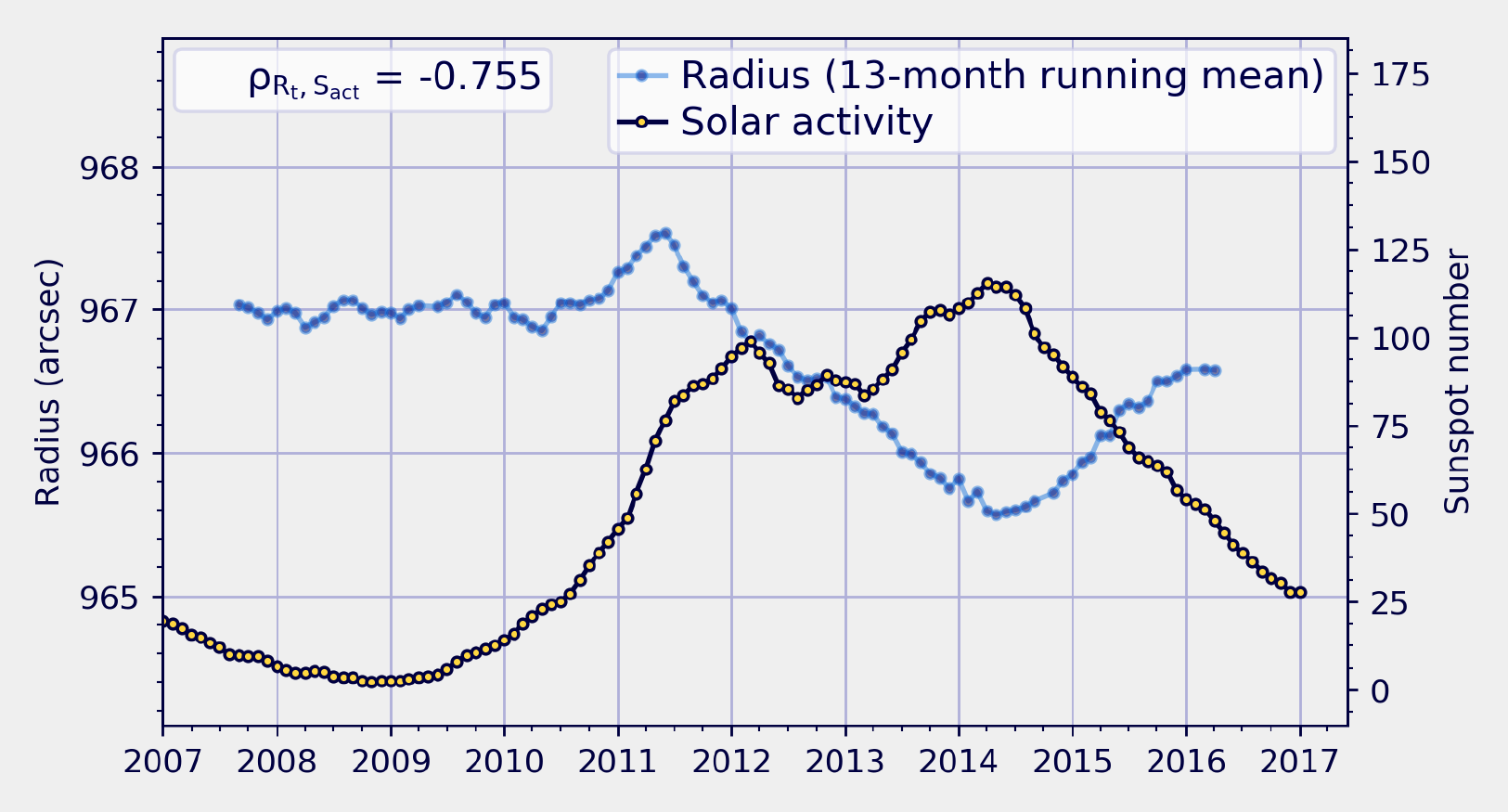}}
			\subfloat
			{\includegraphics[trim={2mm 3mm 2mm 3mm}, clip, height=35mm, width=.5\textwidth ]{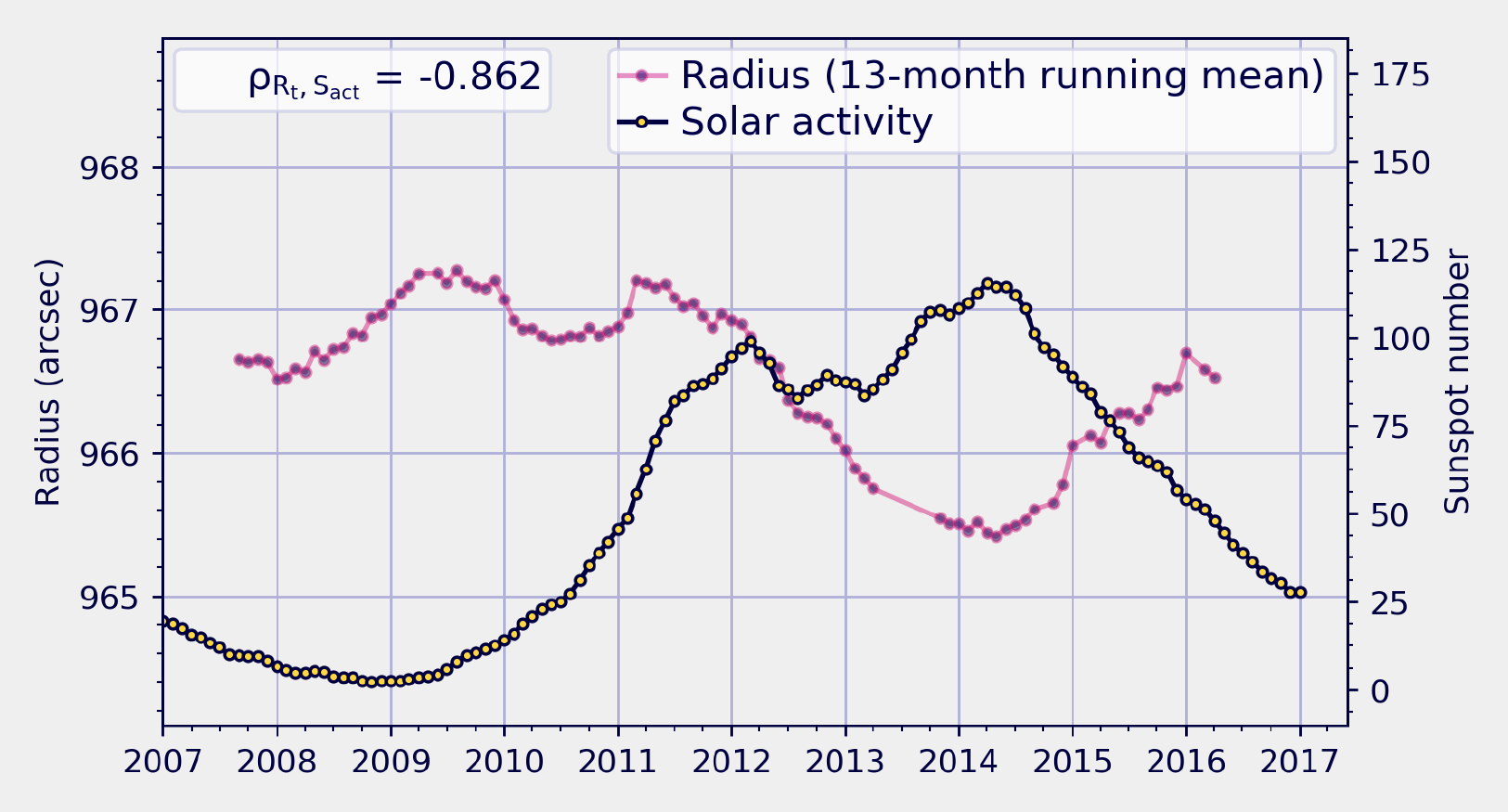}}
			\caption{\small Radius temporal variations (blue and red lines) and sunspot number (black line) from 2007 to 2017 for a 1-month (top) and 13-months (bottom) applied running mean. \textit{Left}: 212 GHz. \textit{Right}: 405 GHz. {\scriptsize Sunspot index number source: SILSO data/image, Royal Observatory of Belgium, Brussels.}}
			\label{correlation}
		\end{figure}

		However, according to \citeonline{kaufmann08}, SST's radiometers were upgraded in 2006 to improve bandwidth, noise, and performance. In 2007, SST's reflector was also repaired providing better antenna efficiency. Thus, we recalculated the mean radius and we checked the PCC of the samples once more, but now from 2007 to 2017. The histograms (mean, median and mode values) remained the same while the recalculated PCC revealed, as shown in Figure \ref{correlation}, for a 1-month running mean, correlation coefficients of $\rho_{0.212 \THz} = -0.493$ and $\rho_{0.405 \THz} = -0.466$, whereas for a 13-month running mean, the coefficients were $\rho_{0.212 \THz} = -0.755$ and $\rho_{0.405 \THz} = -0.853$. These latter values indicate a strong anti-correlation between the 11-year solar activity cycle and the radius temporal variation at THz frequencies.
		
		For a 1-month running mean, the minimum and maximum radii were \ang{;;963.7} and \ang{;;968.9} at 0.212 THz, so a total variation of \ang{;;5.2}, or $3.8 \times 10^6$ m. At 0.405 THz there was a total variation of \ang{;;5.6}, or $4.1 \times 10^6$ m, from \ang{;;962.4} to \ang{;;968.0}. Furthermore, for the 13-month running mean, the 0.212 THz radius varied from \ang{;;965.3} to \ang{;;967.2}, \textit{i. e.}, a variation of \ang{;;1.9}, or $1.4 \times 10^6$ m, whereas at 0.405 THz, the difference was \ang{;;2.1}, from \ang{;;965.0} to \ang{;;967.1}, resulting in an altitude of $1.5 \times 10^6$ m where the subterahertz emissions were roughly produced during solar minimum and maximum.

\section{Conclusions}
\label{conclusion}
	In our study, we analyzed 13,200 maps at 0.212 and 3,423 maps at 0.405 THz -- a total of 16,600 maps approximately. Their radii was determined by fitting the solar limb points, defined at half the quiet Sun value, by a circle. We also analyzed  the solar subterahertz radii variation in time over a decade. The main results are summarized below:
	
	\begin{itemize}
		\item The average radii are \ang{;;966.5} $\pm$ \ang{;;2.8} for 0.2 THz and \ang{;;966.5} $\pm$ \ang{;;2.7} for 0.4 THz;
		\item The heights above the photosphere are $5.0 \pm 2.0 \times 10^3$ km, placing the subTHz emission in the chromosphere; 
		\item There is a strong anti-correlation between the solar activity cycle and the radius variation at both frequencies.
	\end{itemize}
	
	As for the mean subterahertz radius, the heights calculated correspond to the chromospheric layer considering the uncertainty in the measurements. Also, as shown in Figure \ref{prevgraph}, our results fit well within the trend of previous results reported in the literature, where the tendency of the curve seems to get more horizontal at higher frequencies. That could explain why we obtained the same value for the radius at both frequencies. 
	
	Our results on the correlation of radius temporal variations and solar activity cycle are consistent with those of \citeonline{selhorst04} when considering polar measurements. The reason may be related to the existence of polar brightenings that are anti-correlated to the solar magnetic cycle. Another explanation may be that high solar activity is associated with increased magnetic fields, that would lead to a reduction of flux transported by convection, as indicated by \citeonline{gilliland81}.
	
	Our results are crucial to test solar atmospheric models, however more studies of such kind at other wavelengths are needed so that models can be improved.
\section*{Aknowledgements}
	We would like to acknowledge the visionary insight of the late Prof. Pierre Kaufmann who envisioned the many possibilities of monitoring the Sun at high submillimeter-frequencies and build the SST telescope in Argentina. Moreover, we thank CASLEO and CRAAM for the data provided. The authors also thank C. Selhorst, J. Valle and D. Cornejo for fruitful discussions and the anonymous referee for valuable input. F. M. thanks CAPES for the graduate scholarship.
	
	This is a pre-print of an article published in Solar Physics. The final authenticated version is available online at: \url{https://doi.org/10.1007/s11207-017-1216-y}.

\vspace{.6\baselineskip}
\noindent{\small Disclosure of Potential Conflicts of Interest The authors declare that they have no conflicts of interest.}


\postextual

\bibliography{0bibl}

\end{document}